\documentclass[twocolumn,preprintnumbers,amsmath,amssymb,floatfix,superscriptaddress,nofootinbib]{revtex4}
\usepackage{amsmath,hyperref,amssymb}
\usepackage{graphicx}
\usepackage{dcolumn}
\usepackage{bm}
\usepackage{verbatim}
\usepackage{tabularx}
\usepackage{slashed}
\usepackage{cancel}
\usepackage{hyperref}

\def\be{\begin{equation}}
\def\ee{\end{equation}}
\def\bea{\begin{eqnarray}}
\def\eea{\end{eqnarray}}
\def\ba#1\ea{\begin{align}#1\end{align}}
\def\bg#1\eg{\begin{gather}#1\end{gather}}
\def\bm#1\em{\begin{multline}#1\end{multline}}
\def\bmd#1\emd{\begin{multlined}#1\end{multlined}}

\renewcommand{\t}{\tilde}
\newcommand{\sgn}{{\rm{sgn}}}

\begin{document}

\title{Two dimensional non-Fermi liquid metals: a solvable large N limit}

\author{Jeremias Aguilera Damia}
\affiliation{\small \it Centro At\'omico Bariloche, CNEA and CONICET, Bariloche, R8402AGP, Argentina}
\author{Shamit Kachru}
\affiliation{\small \it Stanford Institute for Theoretical Physics, Stanford University, Stanford, CA 94305, USA}
\author{Srinivas Raghu}
\affiliation{\small \it Stanford Institute for Theoretical Physics, Stanford University, Stanford, CA 94305, USA}
\affiliation{\small \it SLAC National Accelerator Laboratory, Menlo Park, CA 94025, USA}
\author{Gonzalo Torroba}
\affiliation{\small \it Centro At\'omico Bariloche and CONICET, R8402AGP, Argentina}

\date{\today}

\begin{abstract}
Significant effort has been devoted to the study of ``non-Fermi liquid" (NFL) metals: gapless conducting systems that lack a  quasiparticle description.   One class of NFL metals  involves a finite density of fermions interacting with soft order parameter fluctuations near a quantum critical point.  The problem has been extensively studied in a large N limit (N corresponding to the number of fermion flavors) where universal behavior can be obtained by solving a set of coupled saddle-point equations.  However a remarkable study by S.-S.~Lee revealed the breakdown of such approximations in two spatial dimensions. We show that an alternate approach, in which the fermions belong to the fundamental representation of a global $SU(N)$ flavor symmetry, while the order parameter fields transform under the adjoint representation (a ``matrix large N" theory), yields a tractable large N limit.   At low energies, the system consists of an overdamped boson with dynamical exponent $z=3$ coupled to a non-Fermi liquid with self energy $\Sigma(\omega) \sim \omega^{2/3}$, consistent with previous  studies.  
\end{abstract}

\maketitle


\section{Introduction}

In many strongly correlated quantum materials, continuous phase transitions into a broken symmetry phase occur at zero temperature as a function of pressure, doping and other non-thermal tuning parameters.  At such a quantum critical point~\cite{Hertz1976}, the metallic fermions scatter off of nearly critical fluctuations of the order parameter, and new universal behavior, inconsistent with Landau's Fermi liquid paradigm, can occur.  Understanding such non-Fermi liquid (NFL) behavior~\cite{Schofield1999,Varma2002} and its relation to high-temperature superconductivity is one of the central challenges of  theoretical physics.   We study a class of quantum critical points that preserve the underlying lattice translational symmetry and are not associated with a conserved order parameter --an example is the Ising nematic transition, which has been observed in several iron-based superconductors~\cite{Shibauchi2013,Kuo2016}, and may play a role in other materials as well~\cite{Stewart1984, Lohneysen2007}.

Near the quantum critical point, only the slowest modes are important; the problem can thus be recast into a quantum field theory involving fermions near the Fermi level coupled to a critical boson (order parameter) by the lowest order interaction allowed by symmetry.  The leading interaction is a Yukawa-type coupling, which is relevant in the renormalization group sense below 3 space dimensions.  As a consequence the theory is strongly coupled in 2 space dimensions, the limit applicable to many quasi-two dimensional quantum materials.  While in recent years numerical methods have revealed a variety of strong coupling effects in two dimensions --for instance via sign problem-free quantum Monte Carlo simulations~\cite{Schattner2016,Berg2019,Xu2019}-- an analytic solution based on a controlled expansion remains elusive.

Given the absence of a perturbative coupling, it is natural to look for a large $N$ expansion to restrict the class of quantum effects that contribute. One possibility is to extend the number of fermion spins from 2 to $N$, and have them interact with a singlet scalar mode; this ``vector large $N$ limit'' has been intensely studied in the literature~\cite{Polchinski1994, Kim1994,Altshuler1994, Rech2006,Lee2009,Metlitski2010,Sur2014,lee2017recent}. However, it was shown in~\cite{Lee2009} that the theory remains strongly coupled due to quantum enhancements at two and higher loops. As a result, the $1/N$ expansion is not enough to make the dynamics tractable. There exist extensions of this limit  that end up being controlled, but this is achieved at the price of adding some new perturbatively small parameter by hand~\cite{Nayak1994, Nayak1994a, Senthil2009, Mross2010, Lee2013}.

In this work we will instead focus on the ``matrix large $N$ limit,'' where $N$ fermion flavors interact with an $N \times N$ matrix-valued boson. This $1/N$ expansion was originally introduced in the context of relativistic quantum field theory, in order to study Yang-Mills theory~\cite{tHooft:1973alw}. It was first applied to NFLs in~\cite{Mahajan2013,FKKRtwo}, and a controlled quantum critical point was shown to arise in an $\epsilon$ expansion around $d=3$ spatial dimensions~\cite{Torroba:2014gqa, Raghu:2015sna}.  We will study this $1/N$ expansion directly in two spatial dimensions and at zero temperature, finding an exactly solvable critical point with non-Fermi liquid behavior.  The exact solution consists of an overdamped order parameter field with dynamical exponent $z_b=3$, coupled to a non-Fermi liquid metal with fermion dynamical exponent $z_f = 3/2$.  Similar solutions have been obtained both in direct perturbation theory~\cite{Rech2006}, and in the vector large N limit.\footnote{See also~\cite{Chakravarty1995, Oganesyan2001} for other methods that give similar self-energy effects.} Here, however, they correspond to a controlled and asymptotically exact solution of an infrared fixed point.  Our results thus provide a controlled framework for understanding non-Fermi liquid behavior. 

The paper is organized as follows. In Sec.~\ref{sec:classical} we present the model and discuss the one-loop QCP. In Sec.~\ref{sec:allN} we extend the validity of the QCP to all orders in the $1/N$ expansion. We do this by determining a low energy limit where the standard large $N$ counting of planar and nonplanar diagrams applies. In Sec.~\ref{sec:concl} we compare with the vector large $N$ expansion, which remains strongly coupled; we track the difference to the qualitatively different behavior of the 't Hooft coupling. We also compare our framework to the holographic approach to non-Fermi liquids, and propose future directions of research. In the Appendices we present an alternative and equivalent renormalization-group analysis, as well as a scaling analysis of a more general model that includes the vector and matrix large $N$ expansions.

\section{The one loop critical point}\label{sec:classical}

Our euclidean action involves a two-dimensional system consisting of fermions ($\psi, \bar \psi$) at finite density interacting with a critical boson $\phi$:
\bea\label{eq:S0}
S&=&\int d\tau d^2x\, \Big \lbrace \frac{1}{2}  {\rm Tr} \left[\frac{1}{c^2} \left( \partial_{\tau} \phi \right)^2 +\left( \nabla \phi \right)^2 \right]\\
&+&\psi^{\dagger}_i \left( \partial_{\tau} + \varepsilon(i \nabla) - \mu_F \right) \psi^i +  \frac{g}{\sqrt{N}} \phi^{i}_{ j} \psi^{\dagger}_i \psi^j  \Big \rbrace \,. \nonumber
\eea
To facilitate an asymptotically exact solution, we impose a 
 global $SU(N)$ flavor symmetry, with $\psi_i, i = 1, \cdots N$ transforming in the fundamental, and $\phi^i_j ,i,j=1 \cdots N$ in the adjoint representation. 
Here, we have tuned to criticality by switching off the boson mass,  $c$ is the boson speed, 
the fermion has a dispersion relation $\varepsilon(\vec k)$ and chemical potential $\mu_F$, and the two fields are coupled via a cubic Yukawa interaction. This is the most relevant interaction consistent with the symmetries, and 
we will show that other interactions, such as the boson $\phi^4$ and the BCS coupling, are irrelevant at the fixed point.

We will first analyze the critical point that arises at one loop, and in Sec. \ref{sec:allN} we will show that all the other corrections vanish in $1/N$. So the fixed point will turn out to be one-loop exact in $1/N$.

The kinematics of the Fermi surface and its coupling to the boson will play an important role in the long distance dynamics. So let us first review the decomposition of fermionic and bosonic momenta. A given point on the one-dimensional Fermi surface is parametrized by the Fermi surface radius $k_F$ and a unit vector $\hat n$. The fermionic momentum is then written as a radial fluctuation \cite{Shankar}, $\vec p = \hat n (k_F + p_\perp)$. 
The Yukawa interaction implies that the boson momentum $\vec q$ behaves as a difference of fermion momenta. Near the point $\hat n$ on the Fermi surface, we will decompose $\vec q = q_\perp \hat n+ \vec q_\parallel$, and will often denote the relative angle by $\cos \theta = \vec q \cdot \hat n /q$.

\begin{figure}[h!]
\centering
\includegraphics[width=.6\linewidth]{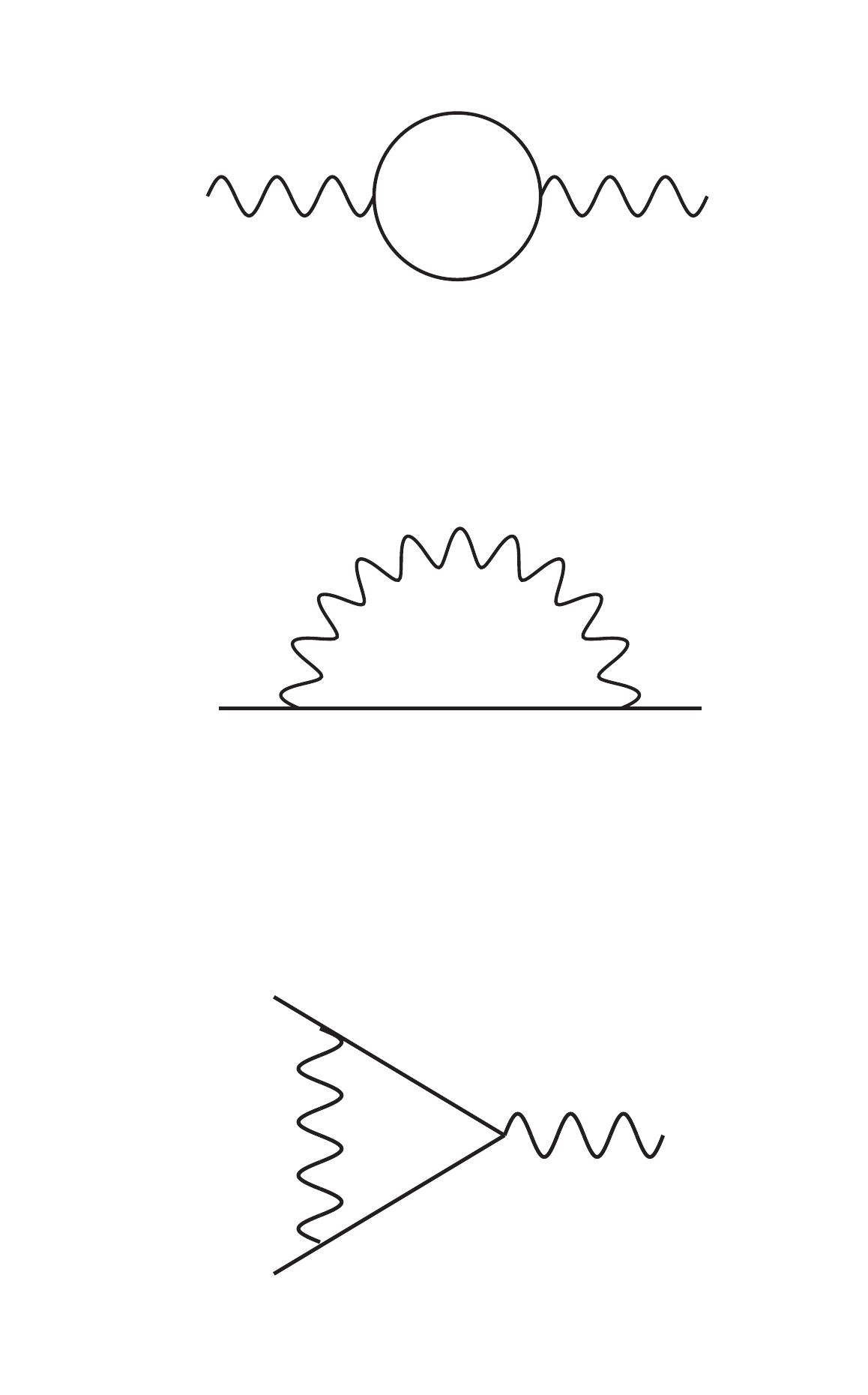}
\caption{\small{One-loop quantum effects: boson self-energy (top), fermion self-energy (middle), and vertex renormalization (bottom). Boson and fermion propagators are denoted by wavy lines and straight lines, respectively.}}\label{fig:oneloop}
\end{figure}

One loop quantum effects induce boson and fermion self-energy corrections; see Fig.~\ref{fig:oneloop}. A standard calculation gives the boson self-energy (Landau damping)
\be\label{eq:LD1}
\Pi(q_0, q) = \frac{k_F}{N} \,\frac{g^2}{2\pi v}\,\frac{|q_0|}{\sqrt{q_0^2+ (v q)^2}}\,.
\ee
While this is a $1/N$ effect, we will include it because it dominates at low energies. 
Including the effects of $\Pi(q_0,q)$, the boson spectral weight dominantly arises from the kinematic regime $|q_0| < vq$, where the characteristic boson speed is slow compared to that of the fermion, and where the boson mixes with the continuum of particle-hole excitations of the Fermi surface.  As a result, the boson gets overdamped, and 
combining (\ref{eq:LD1}) with (\ref{eq:S0}) gives, at low energies, a boson with $z=3$ scaling, $q^3 \sim M_D^2 |q_0|$.  Here we have introduced the Landau damping scale
\be\label{eq:MD}
M_D^2 \equiv \frac{k_F}{N}\,\frac{g^2}{2\pi v^2}\,.
\ee
We will then work with the resummed bosonic propagator \cite{Hertz1976,Torroba:2014gqa} 
\be\label{eq:Dz3}
D^{-1}(q_0, q) \approx q^2+ M_D^2\,\frac{|q_0|}{q}\,,
\ee
which will be shown to be self-consistent.

The computation of the fermion self-energy using this resummed  overdamped boson propagator is standard and results in the following expression:
\be\label{eq:Sigma-1}
\Sigma(p_0) = \frac{g^2}{2\pi \sqrt{3}v}\,\frac{1}{M_D^{2/3}}\,\text{sgn}(p_0)\,|p_0|^{2/3}\,.
\ee
The self-energy is a regular function of momentum, which we have not included here, since it becomes 
irrelevant due to the $z=3$ scaling of the boson internal line. The remaining one loop effect, the vertex correction, is suppressed by $1/N$, analogous to a ``Migdal" approximation in the electron-phonon problem, and can be neglected.

Eqs.~(\ref{eq:Dz3}) and (\ref{eq:Sigma-1}) describe a nontrivial QCP, where the radial fermionic momentum scales differently from the bosonic momentum~\cite{Torroba:2014gqa, Fitzpatrick:2014cfa, Raghu:2015sna}. It is not hard to check that the scale transformations
\be\label{eq:coord-transf}
\omega \to \lambda \omega\;,\;q_\perp \to \lambda^{2/3} q_\perp\;,\;q_\parallel \to \lambda^{1/3} q_\parallel\,,
\ee
and
\be\label{eq:field-transf}
\phi(q_0, q) \to \lambda^{-4/3} \phi(q_0,q)\,,\, \psi(q_0, q) \to \lambda^{-7/6} \psi(q_0, q)
\ee
leave the IR effective action (which includes the above self-energy corrections) invariant. As a result, we obtain a quantum critical point where the boson has scaling dimension and dynamical exponent $(\Delta_\phi = -\frac{4}{3}\,,\,z_b=3)$, and for the fermion, $(\Delta_\psi = -\frac{7}{6}\,,\,z_f=\frac{3}{2})$.\footnote{These are the dimensions in momentum space representation, as in (\ref{eq:field-transf}).}
The only relevant coupling (besides the chemical potential) is the boson mass, which we tune to criticality. The Yukawa interaction becomes marginal at the fixed point, while 4-boson and 4-Fermi interactions are irrelevant. (This is why we neglected them from the beginning). We note that this fixed point agrees with the $\epsilon=1$ limit of the NFL studied in~\cite{Torroba:2014gqa, Raghu:2015sna}, in $d=3-\epsilon$ dimensions. 

Finally, let us determine the energy scale below which we flow to the one-loop QCP. This is the crossover at which the quantum self-energies begin to dominate over the tree-level kinetic terms. This happens when the $z=3$ regime is reached, which requires $q_0^2/c^2 \lesssim \Pi(q_0,q)$ and $q_0^2 \lesssim (vq)^2$. Assuming we are near the mass-shell condition $q^3 \sim M_D^2 q_0$, this gives an energy scale
\be\label{eq:bos-res}
E \lesssim c\frac{\sqrt{k_F g^2}}{N^{1/2}}\,\min(\frac{c^{1/2}}{v},\frac{v^{1/2}}{c})\,.
\ee

\section{Quantum criticality at all orders in $1/N$}\label{sec:allN}

Including the self-energy effects described in the previous section, we
obtain a one-loop QCP with effective Lagrangian 
\be
L_{eff}= L_f+L_b+L_Y
\ee
where
\bea\label{eq:Leff2}
L_f&=& \int dp_\perp\,(k_F d \hat n)\, \psi_{\hat n}^\dag \left(i \beta N^{1/3} \sgn(p_0) |p_0|^{2/3}-v p_\perp\right) \psi_{\hat n}  \nonumber \\
L_b&=& \int dq_\perp dq_\parallel\,\phi \left(q^2 +\frac{\gamma}{N} \frac{|q_0|}{q}\right) \phi \\
L_Y&=& \frac{g}{\sqrt{N}} \int dq_\perp dq_\parallel dp_\perp \,k_F d\hat n\, \phi(q) \psi_{\hat n}^\dag(p+q)\psi_{\hat n}(p)\nonumber\,.
\eea
Here we have introduced the combinations
\be
\beta = \frac{1}{(2\pi)^{2/3} 3^{1/2}} \left(\frac{g^4}{v k_F} \right)^{1/3}\;,\; \gamma = \frac{k_F g^2}{2\pi v^2}\,.
\ee

Using the $1/N$ expansion, we now want to extend this to all loop orders. This, however, encounters some problems due to the fact that the explicit $N$ dependence in the propagators precludes the standard large $N$ counting of planar and non-planar diagrams. In particular, some terms that are irrelevant by the power-counting of (\ref{eq:coord-transf}), are actually enhanced by $N$. A simple example occurs in the bosonic propagator. Here $q_\perp^2$ is irrelevant compared to $q_\parallel^2$, but the $N$-scaling, dictated by the on-shell conditions, is $q_\perp \sim N^{1/3} q_0^{2/3}$, $q_\parallel \sim 1/N^{1/3} q_0^{1/3}$. So $q_\perp^2 \gg q_\parallel^2$ at fixed energy. In other words, the low energy limit does not commute with the large $N$ limit.

We will now argue that the low energy and large $N$ limits can be taken simultaneously, if the external frequencies and momenta scale in a specific way with $N$. To see this, we note that the previous problem --the large $N$ limit ruining the $z=3$ scaling-- is resolved if the low energy limit is taken as $q_0 \sim 1/N^2$. Indeed, this makes $q_\perp$ and $q_\parallel$ above scale with the same power of $N$. Therefore, we will consider the redefinition
\be\label{eq:redef}
p_0 = \frac{1}{N^2} \tilde p_0\,,\, p_\perp = \frac{\beta}{N}\tilde p_\perp\,,\,p_\parallel=\frac{\gamma^{1/3}}{N} \tilde p_\parallel\,.
\ee
We will show that correlation functions with fixed $(\t p_0, \t p_i)$ are described by a QCP that is one-loop exact in the $1/N$ expansion. Before proceeding, we also note that we have introduced factors of $\beta, \gamma$ in (\ref{eq:redef}), so that the engineering dimensions of the new variables, $[\tilde p_0]=1\,,\,[\tilde p_\perp]=2/3\,,\,[\tilde p_\parallel]=1/3$, match the scaling dimensions (\ref{eq:coord-transf}) of the one-loop fixed point. 

The redefinition (\ref{eq:redef}) produces overall powers of $N$ and $(\beta, \gamma)$ in the two-point functions. However, these factors cause no problem, as they can be absorbed into the redefinition of fields.  The canonically normalized fields, where these factors are absorbed, become
\be
\chi_{\hat n} = \frac{\beta k_F^{1/2}}{N^2}  \psi_{\hat n}\;,\;\varphi= \frac{(\beta \gamma)^{1/2}}{N^3}\phi\,.
\ee
Given the engineering dimensions (in Fourier space) $[\psi]=-2\;,\;[\phi]=-5/2$, the dimensions of the canonical fields become $[\chi]=-7/6\;,\;[\varphi]=-4/3$. As expected, these agree with the scaling dimensions (\ref{eq:field-transf}). The last step replaces these redefinitions in the Yukawa coupling; the resulting effective action $S_{eff}= S_f+S_b+S_Y$ reads
\bea\label{eq:qcp}
S_f&=& \int dp_0 dp_\perp\,d \hat n\, \chi_{\hat n}^\dag \left(i  \sgn(p_0) |p_0|^{2/3}-v p_\perp\right) \chi_{\hat n} \nonumber\\
S_b&=& \int dq_0 dq_\perp dq_\parallel\,\varphi \left(q^2 + \frac{|q_0|}{q}\right) \varphi \\
S_Y&=& \frac{ g_*}{\sqrt{N}} \int dq_0 dp_0 dq_\perp dq_\parallel dp_\perp \, d\hat n\, \varphi(q) \chi_{\hat n}^\dag(pq)\chi_{\hat n}(p)\nonumber\,,
\eea
and we have dropped all the tildes from the frequencies and momenta. The coupling evaluates to 
\be\label{eq:gstar}
\frac{g_*^2}{v}=2\pi \sqrt{3}\,.
\ee
This plays the role of the 't Hooft coupling at the fixed point.
In Appendix \ref{app:RG}, we show that the above fixed point action (\ref{eq:qcp}) can equally well be captured by a renormalization group treatment -- see, for instance Eq.~(\ref{eq:fp}).  
Indeed, the scalings and redefinitions that we just performed are automatically included in the RG approach in terms of the running parameters.

Since the fixed point theory has an order one 't Hooft coupling, we expect that we have to resum all planar diagrams that contribute to (\ref{eq:qcp}). Fortunately, they all vanish beyond one loop. This can be seen by noting that planar corrections to the self-energies are resummed in terms of the Schwinger-Dyson equations
\begin{widetext}
\bea\label{eq:SD}
\Pi(q_0,q) &=& \frac{g^2}{N}\int \frac{dk_0}{2\pi}\frac{dk_\perp}{2\pi}\frac{d\theta}{2\pi}\,\frac{1}{i k_0+i \Sigma(k_0)-v k_\perp}\frac{1}{i (k_0+q_0)+i \Sigma(k_0+q_0)-v (k_\perp+q \cos \theta)} \nonumber\\
i \Sigma(p_0) &=&-g^2 \int \frac{dq_0}{2\pi} \frac{q dq}{2\pi} \frac{d\theta}{2\pi}\,\frac{1}{q^2+\Pi(q_0,q)}\,\frac{1}{i q_0+i\Sigma(q_0)- v q \cos \theta}\,.
\eea
\end{widetext}

\begin{figure}[h!]
\centering
\includegraphics[width=1.\linewidth]{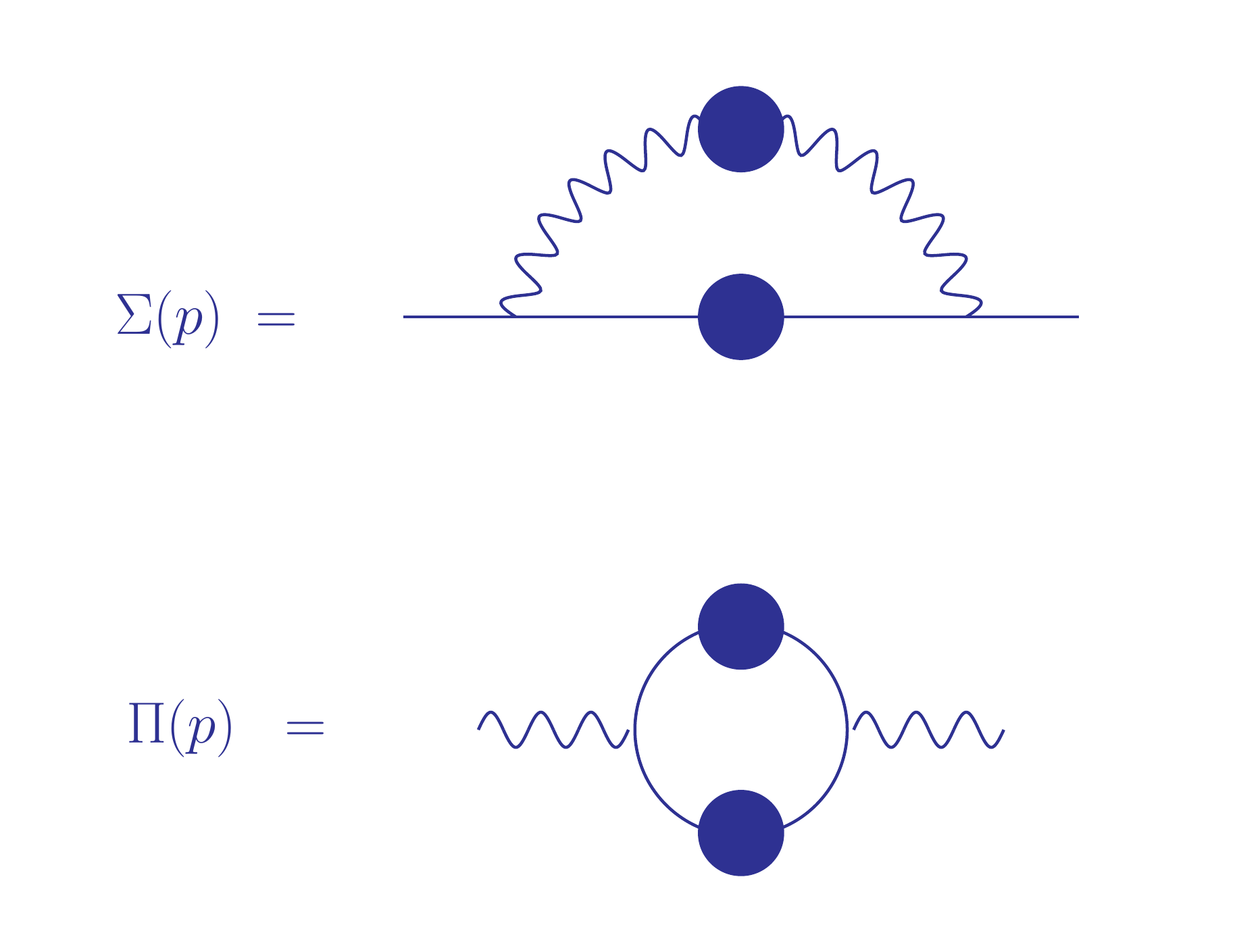}
\caption{\small{Schwinger-Dyson equations for the fermion and boson self-energies, neglecting vertex corrections in the large $N$ theory.}}\label{fig:SDeq}
\end{figure}

Diagrammatically, replacing the one-loop contributions in the right hand side of (\ref{eq:SD}) gives rise to the two-loop rainbow diagrams for $\Pi$ and $\Sigma$, and this continues by induction to higher-loop planar diagrams. This is illustrated diagrammatically in Fig.~\ref{fig:SDeq}. By explicit calculation, (\ref{eq:Dz3}) and (\ref{eq:Sigma-1}) provide a solution to (\ref{eq:SD}), so long as the low energy limit is taken as in (\ref{eq:redef}).\footnote{Above this window, the $z_b =3$ and $z_f =3/2$ scalings are not preserved. } In summary, the one loop result is a self-consistent solution to the Scwhinger-Dyson equations, and all planar contributions beyond one loop vanish in the low energy limit (\ref{eq:redef}).

On the other hand, all nonplanar corrections to the QCP are explicitly suppressed by powers of $1/N$. This can be seen directly from (\ref{eq:qcp}): the usual large $N$ counting of diagrams applies, because $N$ only appears in the cubic interaction and not inside the two-point functions. This is a consequence of the way in which the low energy and large $N$ limits are taken in (\ref{eq:redef}).

Let us also mention that the tree-level irrelevant contributions to the kinetic terms --the boson and fermion frequency terms, and the higher order term $p_\perp^2/k_F$ in the fermion dispersion relation-- can also be seen to be suppressed by powers of $1/N$ compared to the critical terms. The same occurs with terms that have four or more fields in the action. As a result, none of the irrelevant corrections to the QCP are enhanced by powers of $N$. 

We conclude that the one-loop QCP (\ref{eq:qcp}) is exact to all orders in $1/N$. It arises in the simultaneous large $N$ and low energy limit dictated by (\ref{eq:redef}), with $(\tilde p_0, \tilde p_i)$ fixed. This QCP thus provides an example of a solvable non-Fermi liquid in two spatial dimensions.

\section{Discussion and conclusions}\label{sec:concl}

We have shown above that the matrix large $N$ limit provides a controlled set of solutions describing the two dimensional quantum critical metal. This was achieved by taking a simultaneous large $N$ and low energy limit (\ref{eq:redef}). The solvability of the $1/N$ expansion may appear surprising, both from previous results on the vector large $N$ limit~\cite{Lee2009}, and because in general it is very hard to resum the planar expansion in relativistic quantum field theory~\cite{tHooft:1973alw, Maldacena:1997re, Aharony:1999ti}. In order to address this,
let us now briefly discuss the problem from the viewpoint of the renormalization group (RG).  

The self-consistency of the quantum effective action (\ref{eq:qcp}) implies an IR stable RG fixed point.  In App.~\ref{app:RG}, we show how this result can equally well be captured by a renormalization group treatment; see for instance Eq.~(\ref{eq:fp}).  We summarize here the essential features.  In the vicinity of the fixed point, the one-loop beta function for the combination $\alpha \sim g^2/v$ is
\begin{equation}
-\mu \frac{d \alpha}{ d \mu} = c_1 \alpha  - c_2 \alpha^2, 
\end{equation}
where, $\mu$ is the sliding energy scale (the RG flow parameter), and $c_1, c_2$ are positive order one constants.  The first term above describes the tree-level scaling behavior of $\alpha$ at low energies, while the second term contains the effects of quantum self-energy corrections (recall that vertex corrections can be neglected in the large $N$ limit).  As a consequence, there is an IR stable fixed point with an order unity fixed point value $\alpha_* \sim \mathcal O(1)$.  This fixed point precisely corresponds to the action  (\ref{eq:qcp}), where a $z=3$ boson is coupled to a non-Fermi liquid with an order unity 't Hooft coupling (\ref{eq:gstar}).

By contrast, in the vector large $N$ limit, the fermion self-energy is a $1/N$ correction.  As a consequence, the analogous RG flows are described by an equation of the form
\begin{equation}
-\mu \frac{d \alpha}{ d \mu} = c'_1 \alpha  - \frac{c'_2}{N} \alpha^2, 
\end{equation}
and the resulting fixed point value corresponds to $\alpha_* \sim N$.  This theory remains strongly coupled at the purported fixed point and we lose theoretical control.  This is the essence of the problem noted in~\cite{Lee2009}.  We explore this further in App.~\ref{app:redef}, where we construct a scaling theory of the vector large $N$ limit by rescaling momenta, frequency and redefining fields. This rederives an action analogous to Eq.~(\ref{eq:qcp}) with a 't Hooft coupling of order $N$, showing that the theory flows to strong coupling even at leading order in the large $N$ expansion.  

Let us also compare our results with the planar limit of non-abelian gauge theories, and more generally with large $N$ conformal field theories (CFTs). In this case, there is an infinite number of planar diagrams, whose resummation can often be described by a classical gravitational theory in one more dimension~\cite{Maldacena:1997re} (and see e.g.~\cite{Aharony:1999ti} for a review). In contrast, here we have found a finite number of planar diagrams that are ultimately responsible for the QCP. The main difference is that in relativistic theories it is necessary to resum the effect of relevant single-trace interactions of the matrix fields, such as ${\rm tr}(\phi^4)$. This gives rise to an infinite class of planar graphs that contribute. On the other hand, in the nonrelativistic setup of this work, the analog single-trace interactions are irrelevant. This leads to a finite class of diagrams whose effects can be taken into account exactly in the $1/N$ expansion.

In recent years, gauge/gravity duality has provided another framework for obtaining NFLs. See~\cite{Lee:2008xf, Liu:2009dm, Cubrovic:2009ye, Faulkner:2009wj} for some of the original works, and~\cite{Hartnoll:2009sz,Herzog:2009xv,McGreevy:2009xe} for reviews with additional references. These NFLs can be minimally described by coupling a strongly interacting large $N$ CFT to a Fermi surface~\cite{Faulkner:2010tq}. The CFT dresses the Fermi surface into a NFL with self-energy $\sim \omega^{2\Delta-1}$, where $\Delta$ is the dimension of the CFT operator that couples to the fermions. On the other hand, the backreaction of the fermions on the CFT is a negligible $1/N$ effect. Here we find some similarities with our framework, where the $N \times N$ order parameter $\phi$ gives rise to a NFL behavior $\sim \omega^{2/3}$. One important difference, however, is that the dynamics of the $\phi$ field itself is produced by its coupling to the Fermi surface, and does not need to be put in by hand. In any case, the flexibility of these semi-holographic Fermi liquids suggests generalizations of the theory studied in this work, where an overdamped $N \times N$ boson with dynamical exponent $z_b$ is coupled to a Fermi surface with $N$ fermion flavors. We hope to consider this in future work.

To conclude, we have identified a solvable matrix large $N$ limit in which a two dimensional non-Fermi liquid arises at a quantum critical point.  The theory has identical universal power laws to those conjectured in the vector large $N$ theories.  In the future, we wish to study the interplay between non-Fermi liquid behavior and superconductivity in such systems, as well as to study finite temperature thermodynamic and transport properties using the large $N$ expansion.  Lastly, we comment here that while the solvable large $N$ limit provides insights into the nature of quantum materials, it remains unknown whether the precise power laws are the same in realistic systems with $N \sim 1$.  

\acknowledgements

We thank A. Chubukov, L. Delacretaz and M. Zimet for very useful discussions, and J. McGreevy, M. Mulligan, S.S. Lee and H. Wang for comments on the manuscript.
We would like to especially thank Huajia Wang for many fruitful collaborations on non-Fermi liquids.
JAD is supported by CONICET and by a Fulbright - Bunge and Born fellowship.  SK is supported by a Simons Investigator Award and by the NSF under grant PHY-1720397. SR is supported in part by the DOE Office of Basic Energy Sciences, Contract DEAC02-76SF00515.   GT is supported by CONICET (PIP grant 11220150100299), ANPCYT PICT grant 2015-1224, UNCuyo and CNEA.  JAD and GT would like to thank the Stanford Institute for Theoretical Physics for its hospitality during the first stage of this project.

\appendix

\section{Renormalization group analysis}\label{app:RG}

In this Appendix we present a renormalization-group treatment of the QCP discussed in the main text.

It is convenient to introduce the combination
\be\label{eq:alphadef}
\alpha \equiv \frac{1}{6\pi \sqrt{3}}\frac{g^2}{v }\,,
\ee
in terms of which the fermion wavefunction renormalization reads
\be\label{eq:Z1}
Z(p_0)= 1+ \frac{\Sigma(p_0)}{p_0} = 1+ \frac{3\alpha}{M_D^{2/3} |p_0|^{1/3}}\,.
\ee
The RG approach focuses on running (or renormalized) couplings and fields. The renormalized fermion field is related to the original one by $\psi_r = Z^{1/2}  \psi$, which makes the kinetic term canonical. This redefinition gives rise to a running velocity and coupling
\be
v_r(\mu) = Z^{-1}(\mu) v\;,\;g_r(\mu) = Z^{-1}(\mu) g\,.
\ee
Then $g_r(\mu)^2 /v_r(\mu) \sim Z^{-1}(\mu)$. Since this combination scales as $\mu^{1/3}$ in the $z=3$ regime, we will introduce the dimensionless running coupling $\alpha_r(\mu)$ defined as
\be\label{eq:alphadef2}
\alpha = M_D^{2/3} \mu^{1/3} \,Z(\mu)\,\alpha_r(\mu)\,,
\ee
with the power of $M_D$ chosen to match engineering dimensions. Using (\ref{eq:Z1}), we find
\be\label{eq:running-alpha}
\alpha_r(\mu) =\frac{1}{M_D^{2/3}\mu^{1/3}}\,\frac{\alpha}{1+ \frac{3\alpha}{M_D^{2/3}\mu^{1/3}}}\,.
\ee
In the IR, this flows to the fixed-point value
\be\label{eq:fp}
\alpha_r(0) = \frac{1}{3}\,.
\ee

These results can be equivalently obtained from the one-loop beta functions
\bea
\mu \frac{d\alpha_r}{d\mu}&=&-\frac{1}{3}\alpha_r+2 \gamma \alpha_r\,, \nonumber\\
2\gamma &=& - \mu \frac{d\log Z}{d\mu}\,.
\eea
The first term in the coupling beta function is simply the classical scaling dimension at short distance, while the second term is due to the fermion anomalous dimension. The fixed point value is (\ref{eq:fp}), with anomalous dimension $\gamma(0)=1/6$. This gives rise to a quantum kinetic term $\sim \omega^{1-2\gamma}= \omega^{2/3}$.

\subsection{Comparison with the vector large $N$ limit}

Let us compare these results with the vector large $N$ expansion, where $N$ fermion flavors interact with a singlet bosonic mode. We choose a Yukawa coupling $g/\sqrt{N}$, with $g$ fixed at large $N$. 

The first difference with the matrix limit is that Landau damping is enhanced by the fermion flavors running in the loop, and this gives rise to a damping scale $M_D$ independent of $N$. On the other hand, the fermion self-energy is now suppressed by $1/N$, obtaining
\be
\Sigma(p_0) = \frac{g^2}{2\pi \sqrt{3}v\, N}\,\frac{1}{M_D^{2/3}}\,\text{sgn}(p_0)\,|p_0|^{2/3}\,.
\ee
This NFL contribution starts to dominate over the tree-level kinetic term for energies
\be\label{eq:vector-NFL-scale}
E \lesssim \frac{(g^2/v)^3}{M_D^2}\,\frac{1}{N^3}\,.
\ee

Following the RG approach described above gives a beta function for the running coupling
\be
\mu\frac{d\alpha_r}{d\mu}= -\frac{1}{3} \alpha_r+ \frac{1}{N} \alpha_r^2\,,
\ee
and we recall that $\alpha \sim g^2/v$, see (\ref{eq:alphadef}). We then find that the IR fixed point has a 't Hooft coupling $g^2 \sim N$. Since this grows with $N$, we do not expect a well-defined $1/N$ expansion. This is where the crucial difference with the matrix large $N$ limit lies, since the latter has an order one 't Hooft coupling. In Appendix \ref{app:redef}, we explore a general redefinition along the lines of (\ref{eq:redef}) for the vector large $N$ limit, finding agreement with the present RG result.

The growth $g^2 \sim N$ implies a proliferation of higher loop diagrams, and a strongly coupled large $N$ limit. And indeed, this is the main result of~\cite{Lee2009}. While this reference introduced a different genus expansion to take into account this effect, it is currently not known how to resum the leading contributions in the vector large $N$ limit.

\section{Large $N$ field redefinitions}\label{app:redef}

In this Appendix we analyze the field redefinitions required at large $N$, that were used in the main text. This approach will serve to exhibit the qualitatively different behavior of the matrix and vector large $N$ limits.

For generality, we consider a model that can capture both the matrix and vector large $N$ limits:
\begin{widetext}
\bea
S&=&\int dp_0 dp_\perp\,(k_F d \hat n)\, \psi_{\hat n}^\dag \left(i N^x \sgn(p_0) |p_0|^{2/3}-v p_\perp\right) \psi_{\hat n}+\int dq_0 dq_\perp dq_\parallel\,\phi \left(q^2 +N^y \frac{|q_0|}{q}\right) \phi \nonumber\\
&+&\frac{g}{\sqrt{N}} \int dq_0 dp_0 dq_\perp dq_\parallel dp_\perp \,k_F d\hat n\, \phi(q) \psi_{\hat n}^\dag(p+q)\psi_{\hat n}(p, p)\,.
\eea
\end{widetext}
The matrix large $N$ model is recovered for $(x=1/3,y=-1)$, whereas the vector large $N$ corresponds to $(x=-1,y=0)$. We have also set to one dimensionful combinations $\beta, \gamma$ such as those in (\ref{eq:Leff2}).

As in the main text, we seek a redefinition that eliminates the $N$-dependence of the propagators and which preserves the on-shell scalings
\be
v p_\perp \sim N^x |p_0|^{2/3}\;,\; q^3 \sim N^y |q_0|\,.
\ee
For the boson, we do not need in general to redefine $q_\perp$ and $q_\parallel$ by the same factor, but the parallel component should dominate in the kinetic term so that the $z=3$ scaling is maintained. Furthermore, since the bosonic momentum $q_\perp$ adds to the fermion momentum in the Yukawa coupling, $q_\perp$ and $p_\perp$ should be redefined by the same factor. On the other hand, we do not scale $\hat n$, which is the position on the Fermi surface.

Let us then look for a redefinition
\begin{align}
(q_0,p_0)=N^{a}(\tilde{q}_0,\tilde{p}_0) \, , \, (q_\perp&,p_\perp)=N^{b}(\tilde{q}_\perp,\tilde{p}_\perp) \,, \, q_\parallel=N^{c}\tilde{q}_\parallel \nonumber \\ 
\phi=N^{\rho}\varphi \,\,\, &,\,\,\, \psi = N^{\eta}\chi \,.
\end{align}
Preservation of the $z=3$ scaling at large $N$ requires
\be
c\leq b \,\,\, \Rightarrow \,\,\, \tilde{q}^2 \approx \tilde{q}_\parallel^2\,.
\label{z3}
\ee
Homogeneous $N$-scaling of both bosonic and fermionic kinetic terms implies
\be
a=\frac32(b-x) \,\,\, , \,\,\, c=\frac13(a+y)\,,
\ee 
for which condition \eqref{z3} reads
\be
b\leq-x+\frac23 y = \left\lbrace \begin{array}{ccc} -1 & , & x=1/3 \,\, , \,\, y=-1 \\ 1& , & x=-1 \,\, , \,\, y=0 \end{array} \right. 
\ee
For the large $N$ matrix model in the main text we adopted $b=-1$, which gives the largest energy window where the fixed point is valid.
 
Given this and imposing $N$-independence of both bosonic and fermionic kinetic terms determines the field redefinitions
\be
\rho = -\frac12(4b-3x+y) \,\,\, , \,\,\, \eta=-\frac14(7b-3x)\,.
\ee
The resulting action has canonically-normalized two-point functions, and a cubic interaction proportional to $(g/\sqrt{N})N^{-\frac{x}{2}-\frac{y}{6}}$. Therefore,
\be
\lambda = \frac{g^2}{N^{x+\frac{y}{3}}}\,,
\ee
plays the role of the 't Hooft coupling, 
which measures the strength of the coupling in the large N limit. Note the above expression is independent of our choice of $b$ that determines how the external frequencies and momenta scale to zero. 

In the matrix large $N$, with $(x=1/3,y=-1)$, the 't Hooft coupling becomes $\lambda=g^2$, independent of $N$. So there is a well-defined $1/N$ expansion. In contrast, in the vector large $N$ theory we have $(x = -1, y=0)$, and the strength of the coupling is 
\be\label{eq:vectorN}
\lambda = g^2 N\,.
\ee
So we do not expect a well-defined large $N$ expansion, in agreement with~\cite{Lee2009}. In this way, we find a qualitative difference in the matrix and vector large $N$ expansions.

\bibliography{NFL.bib}{}
\bibliographystyle{utphys}
\end{document}